\documentclass[acmsmall, preprint]{acmart}

\setcopyright{cc}
\setcctype{by}
\acmDOI{10.1145/3715737}
\acmYear{2025}
\acmJournal{PACMSE}
\acmVolume{2}
\acmNumber{FSE}
\acmArticle{FSE022}
\acmMonth{7}
\received{2024-09-13}
\received[accepted]{2025-01-14}

\usepackage{subcaption}
\usepackage{multirow}
\usepackage{colortbl}
\usepackage{svg}
\usepackage{microtype}
\usepackage{url}
\usepackage{tcolorbox}
\usepackage{algorithm}
\usepackage{algpseudocode}
\usepackage{geometry}
\usepackage{booktabs}
\usepackage{adjustbox}
\usepackage{tabularx}
\usepackage{soul}

\usepackage{macros}
\newlength{\textfloatsepsave}
\usepackage{listings}
\usepackage{xcolor}

\colorlet{punctcolor}{red!60!black}
\colorlet{desccolor}{green!60!black}
\colorlet{altcolor}{blue!60!black}
\colorlet{kwcolor}{teal!60!black}
\definecolor{keywordcolor}{rgb}{0.13, 0.29, 0.53}

\begin{document}
\title{LlamaRestTest: Effective REST API Testing with Small Language Models}

\author{Myeongsoo Kim}
\orcid{0000-0002-5018-5280}
\affiliation{%
  \institution{Georgia Institute of Technology}
  \city{Atlanta}
  \country{USA}
}
\additionalaffiliation{%
  \institution{AWS, USA; this work was performed before the author joined AWS}
}
\email{wardballoon@gatech.edu}

\author{Saurabh Sinha}
\orcid{0000-0003-4092-2643}
\affiliation{%
  \institution{IBM Research}
  \city{Yorktown Heights}
  \country{USA}
}
\email{sinhas@us.ibm.com}

\author{Alessandro Orso}
\orcid{0000-0003-4516-9320}
\affiliation{%
  \institution{Georgia Institute of Technology}
  \city{Atlanta}
  \country{USA}
}
\email{orso@cc.gatech.edu}

\begin{abstract}
Modern web services rely heavily on REST APIs, typically documented using the OpenAPI specification. The widespread adoption of this standard has resulted in the development of many black-box testing tools that generate tests based on OpenAPI specifications. Although Large Language Models (LLMs) have shown promising test-generation abilities, their application to REST API testing remains mostly unexplored. We present LlamaRestTest, a novel approach that employs two custom LLMs---created by fine-tuning and quantizing the Llama3-8B model using mined datasets of REST API example values and inter-parameter dependencies---to generate realistic test inputs and uncover inter-parameter dependencies during the testing process by analyzing server responses. We evaluated LlamaRestTest on 12 real-world services (including popular services such as Spotify), comparing it against RESTGPT, a GPT-powered specification-enhancement tool, as well as several state-of-the-art REST API testing tools, including RESTler, MoRest, EvoMaster, and ARAT-RL. Our results demonstrate that fine-tuning enables smaller models to outperform much larger models in detecting actionable parameter-dependency rules and generating valid inputs for REST API testing. We also evaluated different tool configurations, ranging from the base Llama3-8B model to fine-tuned versions, and explored multiple quantization techniques, including 2-bit, 4-bit, and 8-bit integer formats. Our study shows that small language models can perform as well as, or better than, large language models in REST API testing, balancing effectiveness and efficiency. Furthermore, LlamaRestTest outperforms state-of-the-art REST API testing tools in code coverage achieved and internal server errors identified, even when those tools use RESTGPT-enhanced specifications. Finally, through an ablation study, we show that each component of LlamaRestTest contributes to its overall performance.
\end{abstract}

\ccsdesc[500]{Software and its engineering~Software testing and debugging}

\keywords{Language Models for Testing, Automated REST API Testing}

\maketitle



\section{Introduction}
\label{sec:introduction}

Web services often rely on Representational State Transfer (REST) APIs for efficient communication and data exchange. REST, a paradigm established by Roy Fielding~\cite{fielding2000architectural}, emphasizes scalability and flexibility, enabling seamless interactions between diverse software systems through standard web protocols~\cite{pautasso2008}. REST APIs are pivotal in modern software architecture, facilitating operations over the internet without requiring the client and the server to know details about each other beyond the shared understanding communicated through the API itself~\cite{richardson2013restful}. Given their widespread adoption and critical role in modern web services, ensuring proper documentation and testing of REST APIs has become increasingly important.

Platforms like APIs Guru~\cite{apis_guru} and Rapid API~\cite{rapidapi} highlight the prevalence of REST APIs by hosting a vast collection of REST API specifications and highlighting the critical role of clear and standardized documentation. REST APIs are typically documented using structured languages such as OpenAPI Specification (OAS)~\cite{openapi} (formerly known as Swagger~\cite{swagger}), RAML~\cite{raml}, and API Blueprint~\cite{apiblueprint}. These specifications provide essential API details, such as available operations, data formats, and response types, offering a structured way of describing REST APIs.

This structured API documentation has led to the development of many automated black-box testing tools for REST APIs (e.g.,~\cite{Corradini2022, atlidakis2019restler, karlsson2020quickrest, martin2021restest, karlsson2020automatic, zac2022schemathesis, wu2022combinatorial, kim2023reinforcement, tcases, liu2022morest, kim2024multi, corradini2024deeprest,10301255}) that take API specifications as input and drive testing from the specifications. Current REST API testing tools employ different techniques, including stateful fuzzing, combinatorial testing, evolutionary algorithms, deep learning, and Q-Learning, to perform effective testing. However, despite advancements in REST API testing, significant challenges remain due to complex input constraints and inter-parameter dependencies among API parameters of real-world RESTful services~\cite{kim2022automated, martin2019catalogue}.

To address these challenges, researchers have turned to recent advances in the field of natural language processing (NLP), leading to several promising approaches for REST API testing. For example, NLP2REST~\cite{kim2023enhancing} employs NLP techniques to extract constraints and dependencies from OAS and convert them into OAS keywords (e.g., \texttt{minimum}, \texttt{maximum}, \texttt{pattern}) that testing tools can use. ARTE~\cite{alonso2022arte} generates realistic parameter inputs by extracting key information from parameter names and descriptions and querying knowledge bases, such as DBPedia~\cite{bizer2009dbpedia}. RESTGPT~\cite{kim2023leveraging} uses Large Language Models (LLMs) to perform rule extraction and parameter input generation. These tools have improved REST API testing effectiveness by extracting additional machine-readable (structured) information from the human-readable portions of API specifications. However, they lack the ability to dynamically interact with testing tools during execution. This limitation prevents them from refining test inputs and rules based on real-time server feedback.

To address these limitations, we present LlamaRestTest, the first black-box testing approach that leverages Language Models (LMs) to incorporate dynamic server feedback during REST API testing. LlamaRestTest consists of two specialized LMs created by fine-tuning the Llama3-8b model~\cite{llama38b}: LlamaREST-IPD for identifying inter-parameter dependencies, and LlamaREST-EX for generating appropriate input values. These models were fine-tuned using a comprehensive dataset combining Martin-Lopez's dependency database~\cite{martin2019catalogue} and over 1.8 million API operation parameters mined from APIs-guru~\cite{apis_guru}. We also created quantized versions of these models to investigate their efficiency-accuracy tradeoff when using lower precision. LlamaRestTest further leverages these models together with reinforcement learning for more effective API exploration. Specifically, we integrated these models with ARAT-RL~\cite{kim2023reinforcement}\footnote{We chose ARAT-RL as its exploration strategy has been shown to be effective among the recent tools~\cite{kim2023reinforcement}.}, a reinforcement learning-based testing approach, enabling dynamic analysis of parameter descriptions and server responses to generate valid parameter combinations and input values during testing. 

To assess the performance of LlamaRestTest, we collected 12 real-world RESTful services, previously evaluated in~\cite{kim2023enhancing,kim2023reinforcement}, including popular services such as Spotify. We compared our tool with RESTGPT to evaluate the effectiveness of LlamaREST-IPD and LlamaREST-EX in identifying inter-parameter dependencies and computing valid input values, respectively, using various configurations, including the base Llama3-8B model, the fine-tuned model, and the fine-tuned model with 2-bit, 4-bit, and 8-bit quantization. Additionally, we investigated how fine-tuning and quantization impact overall testing performance. We then compared LlamaRestTest with four state-of-the-art REST testing tools: RESTler~\cite{atlidakis2019restler}, EvoMaster~\cite{arcuri2019restful}, MoRest~\cite{liu2022morest}, and ARAT-RL~\cite{kim2023reinforcement}. We measured effectiveness using three key metrics widely adopted in API testing research~\cite{golmohammadi2023survey,kim2022automated}: code coverage (for open-source services), operation coverage (for online services), and the ability to detect internal server errors (for all services). Finally, we conducted an ablation study to assess how specification information, server response messages, parameter input generation, and parameter dependency identification each contribute to LlamaRestTest's overall performance.

Our empirical evaluation demonstrated LlamaRestTest's effectiveness across multiple dimensions. In parameter handling, our fine-tuned LlamaREST-EX model achieved a 72.44\% success rate in generating valid inputs, outperforming the base Llama3-8B model (22.94\%) while also being more effective than RESTGPT (68.82\%). For inter-parameter dependencies, LlamaREST-IPD successfully identified 12 out of the 17 dependencies, whereas RESTGPT and the base Llama model identified nine and two dependencies, respectively. In terms of code coverage, the fine-tuned model with 8-bit quantization achieved the best balance, reaching the highest average coverage (55.8\% method, 28.3\% branch, and 55.3\% line). This model processed 37.5 operations per run—14.5 more operations processed per run than the base Llama model (which processed 23 operations). Fine-tuning alone provided modest benefits (50.1\% method, 23.5\% branch, 49.9\% line coverage, and 24.5 operations processed), showing improvements of 0.9 percentage points, 0.3 percentage points, 1.4 percentage points, and 1.5 additional operations respectively over the base model (49.2\%, 23.2\%, 48.5\%, and 23 operations). The 4-bit quantized model maintained strong performance (54.0\% method, 27.9\% branch, 53.5\% line coverage, and 34.4 operations processed), whereas the 2-bit model showed lower but still meaningful improvements (52.9\% method, 25.7\% branch, 52.6\% line coverage, and 25.4 operations processed) compared to both the base and the fine-tuned model, highlighting the benefit of quantization in REST API testing.

When compared against state-of-the-art testing tools (EvoMaster, ARAT-RL, RESTler, and MoRest), LlamaRestTest demonstrated substantial improvements in coverage (by 10.0--22.7 percentage points for method coverage, 9.2--18.7 percentage points for branch coverage, and 10.0--22.1 percentage points for line coverage) while identifying more internal server errors (204 compared to 130--160).
Our ablation study revealed that removing LlamaREST-IPD had the most significant impact, reducing coverage by 6.2--9.2 percentage points, followed by server response analysis (4.4--6.3 percentage points reduction), parameter descriptions (1.4--5.9 percentage points reduction), and LlamaREST-EX (3.0--4.5 percentage points reduction).

The main contributions of this work are:
\begin{itemize}
\item LlamaRestTest, the first LM-based REST API testing approach that leverages server feedback along with fine-tuning and quantization techniques to improve both effectiveness and efficiency of REST API testing. 
\item Empirical results that demonstrate the effectiveness of fine-tuning and quantization in REST API testing and show that LlamaRestTest outperforms other state-of-the-art testing tools. 
\item An artifact~\cite{artifact} that includes the LlamaRestTest tool, benchmarking services, and the data used for fine-tuning, along with our empirical study results.
\end{itemize}

\section{Background and Motivating Example}
\label{sec:background}

\begin{figure}[t]
  \centering
  \Description{OpenAPI specification showing the /elements/area endpoint of Ohsome API, including parameters such as bboxes, bcircles, bpolys, filter, format, and time, along with their descriptions and types.}
  \begin{minipage}{.9\linewidth}
    \begin{lstlisting}[
        language=json,
        basicstyle=\tiny,
        belowskip=-0.7\baselineskip
        ]
/elements/area:
  get:
    summary: Area of OSM elements
    operationId: area
    parameters:
      - name: bboxes
        in: query
        description: WGS84 coordinates in the following format: id1:lon1,lat1,lon2,lat2|id2:lon1,lat1,lon2,lat2|... OR lon1,lat1,lon2,lat2|lon1,lat1,lon2,lat2|...; no default value (one boundary parameter must be defined)
        type: string
      - name: bcircles
        in: query
        description: WGS84 coordinates + radius in meter in the following format: id1:lon,lat,r|id2:lon,lat,r|... OR lon,lat,r|lon,lat,r|...; no default value (one boundary parameter must be defined)
        type: string
      - name: bpolys
        in: query
        description: WGS84 coordinates given as a list of coordinate pairs (as for bboxes) or GeoJSON FeatureCollection. The first point has to be the same as the last point and MultiPolygons are only supported in GeoJSON; no default value (one boundary parameter must be defined)
        type: string
      - name: filter
        in: query
        description: Combines several attributive filters, e.g. OSM type, the geometry (simple feature) type, as well as the OSM tag; no default value
        type: string
      - name: format
        in: query
        description: Output format geojson (for /groupBy/boundary resources only), csv, or json;
        type: string
      - name: time
        in: query
        description: ISO-8601 conform timestring(s); default: latest timestamp within dataset
        type: string
    responses:
      '200':
        description: OK
    \end{lstlisting}
  \end{minipage}
  \caption{A Part of Ohsome API's OpenAPI Specification.}
  \label{fig:ohsome_oas}
\end{figure}

\subsection{REST APIs and OpenAPI Specification}

REST APIs are web APIs that adhere to RESTful principles, enabling data exchange between clients and servers using protocols like HTTP~\cite{berners1996hypertext}. Key concepts of REST include statelessness, where each request is independent; cacheability, allowing response storage; and a uniform interface that simplifies client-server interactions~\cite{fielding2000architectural, richardson2013restful}. Clients interact with web resources via HTTP requests, performing operations such as creating, reading, updating, or deleting data using methods like POST, GET, PUT, and DELETE. Each HTTP method paired with a resource endpoint defines an operation, with requests containing headers, payloads, and metadata. Responses consist of headers, bodies, and status codes indicating success (2xx), client errors (4xx), or server errors (500).

To effectively document and standardize these REST API interactions, the software industry has widely adopted formal specification formats. OAS~\cite{openapi} is a crucial standard in the realm of RESTful API design and documentation. As an industry-standard format, it delineates the structure, functionalities, and anticipated behaviors of APIs in a standardized and human-accessible way. Figure~\ref{fig:ohsome_oas} presents a portion of the Ohsome API's OAS and illustrates the structured approach of defining API operations in OAS. For instance, it specifies a GET request at the \texttt{\small /elements/area} endpoint, aimed at calculating the area of OpenStreetMap (OSM) elements. This operation contains multiple query parameters, such as \texttt{\small bboxes}, \texttt{\small bcircles}, and \texttt{\small bpolys}, each requiring specific coordinate formats and types. Parameters like \texttt{\small filter}, \texttt{\small format}, and \texttt{\small time} are also included, offering additional functionality and flexibility in data querying. The API's response to a successful query typically includes a \texttt{2xx} status code, indicating the success of the operation.


\subsection{REST API Testing and Tools for Enhancing OpenAPI Specifications}
REST API testing is a critical practice for ensuring the proper functioning of web services \cite{saleem2016quality}. Leveraging REST API documents such as OAS, numerous black-box REST API testing tools have emerged \cite{Corradini2022, atlidakis2019restler, karlsson2020quickrest, martin2021restest, karlsson2020automatic, zac2022schemathesis, wu2022combinatorial, kim2023reinforcement, tcases, liu2022morest, stennett2025autoresttest, corradini2024deeprest}. These tools aim to detect issues like internal server errors (e.g., HTTP 500 status code) by systematically exploring API endpoints and their parameters. While they primarily rely on machine-readable parts of API specifications \cite{kim2022automated, kim2023enhancing}, these tools often fall short in complex scenarios, particularly when handling specific data formats or inter-parameter dependencies (IPDs). For instance, when testing the Ohsome API, they may generate random string values for parameters like \texttt{bboxes}, \texttt{bcircles}, and \texttt{bpolys}, failing to account for their specific constraints. 

To address these limitations, assistant tools such as NLP2REST \cite{kim2023enhancing} and RESTGPT \cite{kim2023leveraging} have emerged to enhance REST API documents by leveraging natural language descriptions. These tools augment OAS with machine-readable rules, including type/format specifications, constraints, example values, and IPDs. IPDs, as studied by Martin-Lopez et al.~\cite{martin2019catalogue}, are particularly important as they define relationships between API parameters where the value of one parameter affects another, ensuring valid and contextually appropriate parameter combinations for testing. Among these enhancement aspects, we particularly focus on example values and IPDs because the type/format and constraints rules are ultimately aimed at generating valid parameters for REST API testing.

\subsection{Large Language Models}

LLMs, such as the Generative Pre-trained Transformer (GPT) series, are at the forefront of advancements in NLP~\cite{hadi2023large}. A language model essentially understands, interprets, and generates text in a manner similar to how humans use language \cite{openai2023gpt4}. These models are trained on extensive collections of text, enabling them to learn a wide range of linguistic patterns and styles. The GPT series, including well-known models such as GPT-3, exemplifies these advanced LLMs \cite{brown2020language}. Trained on diverse and vast datasets, they have demonstrated promising ability in producing text that is strikingly human-like, making them valuable for a variety of applications. This has expanded the applicability of NLP to challenging areas ranging from education to customer service \cite{baidoo2023education,cotton2023chatting}. Recently, LLMs have also been applied for automating software testing tasks, showing promising results in test case generation and bug detection (e.g.,~\cite{10298372,codamosa,pan2025asternaturalmultilanguageunit,alrashedy2024languagemodelsbetterbug}).

Quantization is a technique used to reduce the computational resources required while minimizing the loss in model performance~\cite{jacob2018quantization}. It involves converting standard floating-point representations of a model's weights and activations into lower-precision formats, such as 8-bit, 4-bit, or even lower~\cite{dettmers2023case,li2024norm}. The primary advantage of quantization is the significant reduction in memory usage and computational demand~\cite{ali2024memory}. This technique is particularly relevant in deploying LLMs, which are known for their massive parameter counts, often reaching hundreds of billions and requiring extensive computational resources. By applying quantization, LLMs can be made more accessible for real-world applications, enabling their deployment on devices with limited resources while maintaining acceptable performance levels. For instance, quantizing a 175B parameter model from 16-bit to 4-bit precision can reduce its memory footprint by up to 75\%, making it feasible to be run on consumer-grade hardware~\cite{dettmers2023case}.

Fine-tuning is a technique for adapting pre-trained LLMs for specific tasks, allowing them to perform better on specialized datasets by adjusting their learned representations~\cite{brown2020language}. A notable advancement in this area is the introduction of Low-Rank Adaptation (LoRA) techniques, which modify only a small set of model parameters during training, specifically the attention mechanism's low-rank components. This approach minimizes the computational overhead and reduces the number of parameters updated, facilitating efficient fine-tuning of large models~\cite{hu2021lora}. Building on the principles of LoRA, Quantized Low-Rank Adaptation (QLoRA) incorporates quantization into the low-rank adaptation process to further decrease the memory and computational demands. By reducing the precision of the low-rank factors, QLoRA maintains performance while enhancing the model's efficiency, making it ideal for deployment in resource-constrained environments~\cite{dettmers2024qlora}.



\subsection{Motivating Example}

REST APIs with complex parameter formats and IPDs present significant testing challenges. The Ohsome API, which calculates the area of OpenStreetMap (OSM) elements, exemplifies such complexity. Consider the \texttt{\small /elements/area} operation of the API (Figure~\ref{fig:ohsome_oas}). According to the specification, this operation requires one of three mutually exclusive boundary parameters---\texttt{\small bboxes}, \texttt{\small bcircles}, or \texttt{\small bpolys}---each with specific and intricate formatting rules. For instance, \texttt{\small bboxes} requires WGS84 coordinates in the format ``lon1,lat1,lon2,lat2'', whereas \texttt{\small bcircles} requires coordinates along with a radius in meters. The API's complexity stems from multiple parameters: \texttt{\small filter} (which combines OSM type, geometry type, and tag filters), \texttt{\small format} (which specifies output as geojson, csv, or json), \texttt{\small time} (which accepts ISO-8601 timestrings), and various boundary parameters that require specific coordinate formats. 

Current state-of-the-art testing tools often generate invalid API requests, without considering the parameter value formats and the IPDs. For example, these tools might generate requests that include both \texttt{\small bboxes} and \texttt{\small bcircles}, violating the mutual-exclusivity requirement, and produce incorrectly formatted coordinates. Although REST API testing assistant tools like RESTGPT enhance the specification by providing more detailed descriptions---such as the correct WGS84 coordinate format---they still struggle with understanding and enforcing IPDs. For instance, RESTGPT fails to recognize the mutual exclusivity of \texttt{\small bboxes}, \texttt{\small bcircles}, and \texttt{\small bpolys}, and it does not provide valid values for the \texttt{\small filter} parameter, which must adhere to the OSM tagging schema (e.g., ``amenity=school'' or ``building=residential''). These issues reduce test coverage and effectiveness.

LlamaRestTest addresses these challenges with an adaptive two-stage approach involving LlamaREST-EX and LlamaREST-IPD, both fine-tuned for REST API testing purposes. Initially, LlamaRestTest generates inputs randomly. If the request fails, LlamaRestTest uses LlamaREST-IPD and LlamaREST-EX to identify valid parameter sets and values. For example, if the initial request does not specify any boundary parameters, the response may return an error message such as: ``\emph{The query did not specify any parameter. Please remember: You need to define one of the boundary parameters (bboxes, bcircles, bpolys).}''  Similarly, if a request contains more than two boundary parameters, the server would respond with the error message ``\emph{The query should not have more than one boundary parameters (bboxes, bcircles, bpolys).}''

LlamaREST-IPD processes the operation’s parameter descriptions along with the server's response to identify parameter dependencies. In this case, it detects from the server message that only one boundary parameter can be defined at a time. Using this feedback, LlamaREST-IPD refines the next request, ensuring that only one valid boundary parameter (e.g., \texttt{\small bboxes}) is included. LlamaREST-EX then enhances this process by generating semantically valid values for the parameters, such as correctly formatted \texttt{\small bboxes} values (``lon1,lat1,lon2,lat2'') or appropriate OSM tags for the \texttt{\small filter} parameter. By combining the adaptive learning from LlamaREST-IPD with the input-generation capabilities of LlamaREST-EX, LlamaRestTest can effectively create valid parameter sets that comply with the API's parameter constraints and formatting rules. 

While the effectiveness of LlamaREST-IPD and LlamaREST-EX is important, we also explore the benefit of efficiency through quantization. This is important because LlamaRestTest can perform extensive analysis of server responses---involving LLM inferencing---to refine its test cases adaptively. By applying quantization, we reduce the computational load and inference time, enabling LlamaRestTest to handle these frequent interactions with lower overhead. Our goal is to balance effectiveness (through fine-tuning) and efficiency (via quantization).

\begin{figure}[t]
\centering
\Description{A diagram showing the overview of our approach.}
\includegraphics[width=\columnwidth]{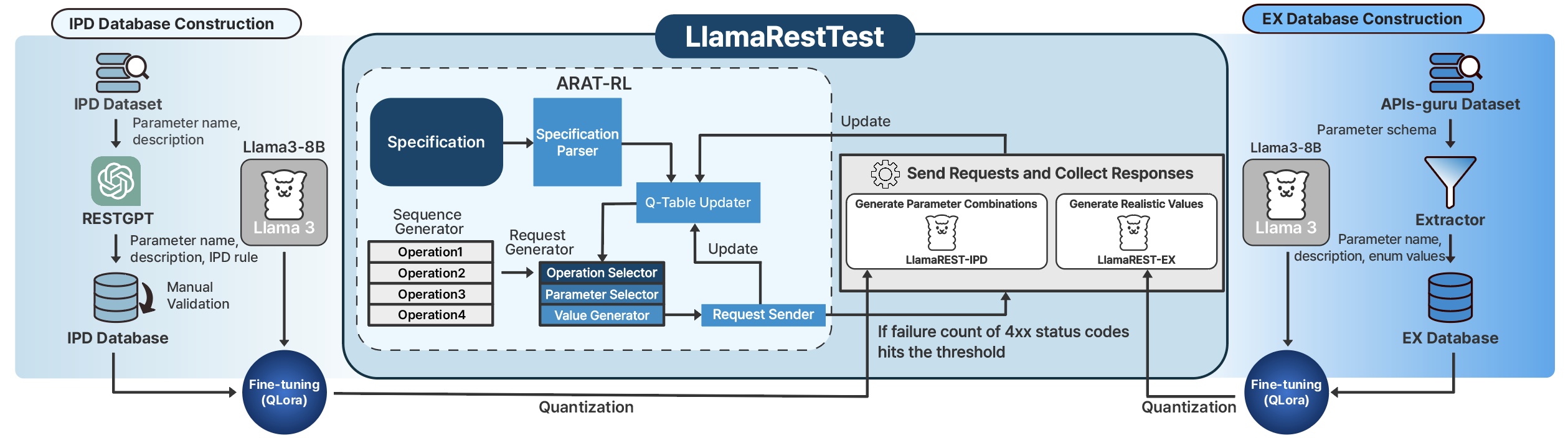}
\vspace{-25pt}
\caption{Overview of our approach.}
\label{fig:overview}
\end{figure}

\section{Our Approach}
\label{sec:our_approach}

Figure~\ref{fig:overview} presents an overview of LlamaRestTest. LlamaRestTest is designed to address the challenges of REST API testing, particularly in handling complex IPDs and generating valid inputs. In this section, we discuss the key features of LlamaRestTest, including model selection, database construction, fine-tuning, and quantization, which together enable adaptive, efficient, and effective testing. These components work in synergy to generate and refine API requests dynamically based on server feedback, improving both test coverage and fault detection. 

\subsection{Base Model Selection}
LlamaRestTest uses LLMs to identify IPDs and generate valid values for API parameters. The closest approach is RESTGPT which leverages OpenAI's ChatGPT-3.5 Turbo model to enhance OpenAPI specifications with rules for REST API testing, perform similar tasks---i.e., identifying IPDs and generating example values. However, the cost---approximately \$10 per specification---poses a challenge, especially when considering the large number of LLM interactions associated with dynamic server interactions. Moreover, RESTGPT focuses solely on static specifications, whereas our approach seeks to leverage server response messages at runtime. Given also the potential exposure of sensitive information in REST API testing scenarios, we prioritized using open-source models that can be hosted on the user's computing resource.

Our goal was to employ powerful yet lightweight LLMs to reduce costs, improve speed, and ensure compatibility with various computing environments, including CPU-based systems. However, we encountered challenges with LLMs, such as the Llama3-8B model, which still requires considerable GPU resources \cite{llama2gpu}. Additionally, Llama3-8b lacked the capability to generate valid IPD rules and example values tailored for REST APIs when tested using prompts from RESTGPT, as shown by our results  (Table~\ref{tab:llamarestex} and Table~\ref{tab:llamarestipd}). 

To address these issues, we applied fine-tuning to enhance the model’s capabilities for REST API testing and employed quantization techniques to make the model more lightweight. Another reason for selecting Llama3-8B as our baseline model is its architecture's support for various quantization options, including popular 2-bit, 4-bit, and 8-bit integer formats~\cite{frantar2022gptq,yao2021hawq,shen2020q}. Llama3 is also open-source and offers a GPT-Generated Unified Format (GGUF) transformation option---a binary format designed for fast model loading and saving.



\subsection{Database Construction}

Figure~\ref{fig:overview} illustrates how we constructed the fine-tuning databases. We systematically created two databases: the IPD database and the example (EX) database. The goal of this process was to collect parameter names, descriptions, and associated rules for the IPD database, which includes the Requires, Or, OnlyOne, AllOrNone, ZeroOrOne, and Arithmetic/Relational/Complex IPD rules~\cite{martin2019catalogue}. For the EX database, we collected parameter names, descriptions, and semantically valid values.
    
For mining IPD data, we used Martin-Lopez's IPD database \cite{martin2019catalogue}, which contains manually collected dependencies from REST APIs. To ensure a fair evaluation on our benchmark, we excluded any OpenAPI specifications that were included in our benchmark dataset. After filtering, we were left with 551 parameters with IPDs. For creating the EX database, we leveraged APIs-guru~\cite{apis_guru}, which provides a large collection of OAS. We gathered 4,133 OpenAPI specifications from this dataset, excluding those used in our benchmark. These specifications include over 1.8 million parameter sets for REST APIs. We specifically targeted parameters with enumerated values which are predefined set of valid values (e.g., \texttt{json} and \texttt{xml} for a parameter on data format).

To generate IPD rules for the parameters in the IPD database, we used RESTGPT~\cite{kim2023leveraging}. RESTGPT extracts IPD rules from natural language descriptions in API specifications. After extracting IPD rules automatically using RESTGPT, we manually validated the rules. The manual validation involved checking whether the extracted IPD rules accurately represented the intended parameter dependencies based on their descriptions. Out of 551 IPDs, 25 (4.5\%) required manual fixing. These were primarily complex cases involving more than two parameters or compound rules. One of the authors, experienced in REST-API testing, performed the validation. The process took approximately five hours in total.

Rules were validated based on their logical consistency with the parameter descriptions. For example, a rule initially extracted as ``Requires(A, B) AND Requires(B, C)'' might be corrected to ``AllOrNone(A, B, C)'' after manual validation. Similarly, the rule ``OnlyOne(X, Y) OR ZeroOrOne(Z)'' might be modified to ``OnlyOne(X, Y, Z)'' to reflect the intended behavior more accurately. It is important to note that manual validation is a one-time step per API version. The validation step is optional and should have limited effect on the results, especially for our dataset where only 4.5\% of IPDs required correction. To support reproducibility, we have shared our models and the validated training dataset in our artifact~\cite{artifact}.

For the EX database, we focused on extracting parameters with enumerated values. From the API documentation, we systematically extracted the \textit{parameter name}, the corresponding \textit{description}, and the specified \textit{enum values} to create the training dataset. This data was structured such that the parameter name and description served as inputs, whereas the valid enum values were used as outputs for model fine-tuning.

The IPD and EX databases share a common schema to ensure compatibility and ease of use during the fine-tuning process. Each database entry includes the following fields:
\begin{itemize}
    \item \textbf{Parameter Name}: Specifies the name of the API parameter (e.g., \texttt{bboxes}, \texttt{filter}, \texttt{format}).
    \item \textbf{Description}: A brief description of the parameter’s purpose, extracted from the OAS or generated by RESTGPT.
    \item \textbf{Inter-Parameter Dependency (IPD)}: For the IPD database, this field outlines the dependency rules associated with the parameter (e.g., mutually exclusive parameters or required groups of parameters). These rules were generated by RESTGPT and manually validated.
    \item \textbf{Example Values (EX)}: For the EX database, this field contains an array of valid enum values for the parameter, as specified in the OpenAPI documentation.
\end{itemize}


\subsection{Fine-Tuning and Quantization}

\subsubsection{Dataset Preparation}

The fine-tuning datasets for IPD detection and example value generation were preprocessed and reformatted into a custom tokenized format suitable for fine-tuning. Initially, we attempted to fine-tune a single model for both tasks. However, our preliminary investigations revealed that this approach led to a poor performance in IPD detection, with the model displaying catastrophic forgetting issues. This led us to the decision to create separate fine-tuned models to improve overall performance.

For each parameter, we used a structured format with special tokens to provide context and guide the model's training. The LlamaREST-IPD model's input sequences follow this structure:

\begin{footnotesize}
\begin{verbatim}
<s>[INST] Find Inter-Parameter Dependency for the parameter below
name: travelMode
description: If startTime is present, travelMode must be 'driving'. [/INST]
IF startTime THEN travelMode == 'driving';</s>
\end{verbatim}
\end{footnotesize}

This format includes the parameter’s name and description, followed by the dependency rule in a structured manner, enclosed with special tokens \texttt{\small <s>} and \texttt{\small [INST]}. Here, \texttt{\small <s>} and \texttt{\small </s>} mark the start and end of the prompt, respectively. Similarly, \texttt{\small [INST]} and \texttt{\small [/INST]} indicate the start and end of the instruction. The text between \texttt{\small <s>[INST]} and \texttt{\small [/INST]} is the instruction, which provides the context; the text after \texttt{\small [/INST]} presents the expected rule. This setup allows the model to learn both the natural language description and the corresponding logic for IPDs.

For LlamaREST-EX, the data is formatted to extract and list valid example values for each parameter. An example of this structure is:

\begin{footnotesize}
\begin{verbatim}
<s>[INST] Find example values for the parameter below in a list format 
name: Content-Type 
description: The content type. [/INST] 
['application/x-www-form-urlencoded', 'application/json', ...]</s>
\end{verbatim}
\end{footnotesize}

This tokenization ensures that the model understands the task of generating example values for each parameter in list format. The input includes the parameter name and description, while the output provides a list of valid example values. We experimented with different tokenization schemes and ultimately chose Llama2 tokens, as Llama3's agentic token support is not required for REST API testing tasks, and Llama2 tokens are simpler. The structured format and custom tokens facilitate efficient parsing of IPDs and example values during fine-tuning.

\subsubsection{Parameter-Efficient Fine-Tuning}
Fine-tuning plays a pivotal role in enhancing the performance of LlamaREST-IPD and LlamaREST-EX by enabling the models to adapt to the tasks of IPD detection and example value generation. Fine-tuning can be performed in two ways: Full Parameter Fine-Tuning (FFT) and Parameter-Efficient Fine-Tuning (PEFT). FFT requires large datasets and extensive computational resources, whereas PEFT is designed to work efficiently with smaller datasets, making it a better fit for our application. Our IPD dataset contains only 551 parameters, which is insufficient for FFT, as prior research has shown that fewer than 1,000 examples do not yield significant performance improvements with FFT~\cite{vieira2024much}. We therefore employed PEFT, which optimizes the model by fine-tuning only a subset of its parameters. This selective updating process allows it to efficiently adapt to new tasks using smaller datasets and fewer computational resources.

The two most commonly used PEFT methods are LoRA~\cite{hu2021lora} and QLoRA~\cite{dettmers2024qlora}. Both approaches allow fine-tuning with significantly reduced computational requirements. LoRA focuses on injecting trainable low-rank matrices into transformer layers, whereas QLoRA quantizes model weights during training, further reducing memory usage and computational costs. Given the limitations of our compute environment, which includes a Google Colab machine with a 40GB A100 GPU, we opted for QLoRA. QLoRA's quantization compresses the model, letting it fit within our memory constraints while still performing the necessary updates.

We based our fine-tuning configuration on the recommended settings from Meta's official guidelines~\cite{llama3finetuning}, making necessary adjustments to accommodate our hardware. Specifically, we reduced the batch size to 8 to fit within the 40GB VRAM of the A100 GPU. As mentioned, the fine-tuning process leverages a structured dataset specifically designed for IPD detection and value generation tasks, using custom tokens to represent the relationships between parameters. These tokens guide the model’s understanding of the dependencies and example values, allowing for updates of only the relevant parts of the model during fine-tuning. This adaptation allows for faster convergence and lower computational costs compared to full parameter tuning.

To prevent overfitting and reduce unnecessary computation, we applied an early stopping strategy~\cite{dodge2020fine}. Initially, we set the epoch limit to 100 but continuously monitored the training loss. Early stopping was triggered when the loss plateaued—meaning further training would no longer improve the model's performance. This resulted in 35 epochs for LlamaREST-EX and 51 epochs for LlamaREST-IPD. 

\subsubsection{Quantization}

After fine-tuning, we applied quantization to optimize the model for deployment in resource-constrained environments and improve speed. Quantization reduces the model's memory footprint by representing its weights with fewer bits, making it faster and more efficient while maintaining acceptable performance. Using the llama.cpp library, we experimented with three popular quantization levels to balance model size and accuracy: 2-bit, 4-bit, and 8-bit quantization~\cite{frantar2022gptq,yao2021hawq,shen2020q}. Each level presents different trade-offs between memory usage and performance, with 2-bit offering the highest compression and 8-bit maintaining nearly full accuracy.

These quantized versions let us assess the trade-off between effectiveness (accuracy) and efficiency (model size and inference speed) for various deployment scenarios in REST API testing. The 2-bit quantization is ideal for highly memory-constrained devices such as edge environments (e.g., 2.96 GB), 4-bit strikes a balance for mid-tier resources (e.g., 4.58 GB), and 8-bit is suited for relatively high-resource environments where performance is a priority (e.g., 7.95 GB). 
\subsection{REST API Testing}

Figure~\ref{fig:overview} illustrates the LlamaRestTest framework, which integrates the ARAT-RL approach for adaptive REST API testing. A key component of ARAT-RL is the Q-Table, which prioritizes API operations and parameters. The \textit{Q-Table Updater} updates Q-values based on responses received from API requests. Specifically, successful responses (2xx) reduce the Q-value, encouraging exploration of other operations, while failure responses (4xx or 500) increase the Q-value, directing further testing toward problematic areas. This update follows the Q-learning formula:

\begin{equation}
Q(s, a) \leftarrow Q(s, a) + \alpha [r + \gamma \max Q(s', a') - Q(s, a)]
\end{equation}

where \(\alpha\) is the learning rate, \(\gamma\) is the discount factor, and \(r\) is the reward assigned based on the response. As shown in Figure~\ref{fig:overview}, the \textit{Request Sender} interacts with the \textit{Q-Table Updater}, dynamically adjusting priorities to optimize testing coverage.

Once initialized, ARAT-RL applies an $\epsilon$-greedy strategy to balance exploration (testing less-explored operations) and exploitation (focusing on previously promising combinations). The \textit{Operation Selector} chooses an API operation based on its Q-value, followed by the \textit{Parameter Selector}, which assigns probabilities to parameters, with more frequently used parameters being favored. The \textit{Value Generator} produces parameter values, typically chosen randomly but in conformance with the expected types.

When repeated failures occur (e.g., multiple \texttt{4xx} responses), LlamaREST-IPD and LlamaREST-EX are activated. The LlamaREST-IPD model is prompted in the same format employed during fine-tuning:

\begin{footnotesize}
\begin{verbatim}
<s>[INST] Find Inter-Parameter Dependency for the parameter below  
name: {parameter name}  
description: {parameter description and server response message} [/INST]
\end{verbatim}
\end{footnotesize}

This prompt directs the model to analyze parameter descriptions and server response messages, identifying dependencies, such as when a parameter is required only if another parameter is present.

Similarly, LlamaREST-EX generates example values for parameters using the following prompt structure, which was used for fine-tuning:

\begin{footnotesize}
\begin{verbatim}
<s>[INST] Find example values for the parameter below in a list format  
name: {parameter name}  
description: {parameter description and server response message} [/INST]
\end{verbatim}
\end{footnotesize}

This enables LlamaREST-EX to produce valid, realistic parameter values based on the available descriptions and server responses.

After parameters and values are refined, the \textit{Request Sender} submits the API request. Depending on the API response, the \textit{Q-Table Updater} adjusts the priorities of the operation-parameter combinations. Successful requests lower the priority of that combination, encouraging exploration, whereas failures assign positive rewards, driving further probing of problematic parts of the API.

\section{Evaluation}
\label{sec:evaluation}

In this section, we present the results of empirical studies conducted to assess LlamaRestTest, focusing on answering the following research questions:

\begin{enumerate}
\item \textbf{RQ1}: How do LlamaREST-EX and LlamaREST-IPD compare with the state-of-the-art REST API testing assistant tool, RESTGPT, in terms of identifying inter-parameter dependencies and generating valid inputs for REST API testing?
\item \textbf{RQ2}: How do fine-tuning and quantization affect REST API testing in terms of achieved code coverage and operation coverage?
\item \textbf{RQ3}: How does LlamaRestTest compare with state-of-the-art REST API testing tools in terms of achieved code coverage and operation coverage?
\item \textbf{RQ4}: In terms of error detection, how does LlamaRestTest perform in triggering 500 (Internal Server Error) responses compared to state-of-the-art REST API testing tools?
\item \textbf{RQ5}: How do analysis of parameter description, analysis of server response messages, LlamaREST-IPD, and LlamaREST-EX contribute to the overall performance of LlamaRestTest in terms of code coverage and operation coverage?
\end{enumerate}

\begin{table}[t]
    \centering
    \begin{minipage}[t]{.41\linewidth}
        \centering
        \caption{REST services used in the evaluation.}
        \vspace{-7pt}
        \resizebox{.95\columnwidth}{!}{%
        \begin{tabular}{lrr}
            \toprule
            \textbf{REST Service} & \textbf{Lines of Code} & \textbf{Operations} \\ \midrule
            Features-Service & 1688 & 18\\
            Genome-Nexus & 22143 & 23\\ 
            Language-Tool & 113170 & 2 \\ 
            Market & 7945 & 13 \\ 
            Person-Controller & 601 & 12\\ 
            Project-Tracking-System & 3784 & 67\\ 
            Rest-Countries & 1619 & 22\\ 
            User-Management & 2805 & 22\\ 
            YouTube & 2422 & 1\\
            FDIC & -- & 6\\
            Ohsome & -- & 122\\
            Spotify & -- & 12\\
            \bottomrule
        \end{tabular}
        }
        \label{tab:service_lines_of_code}
    \end{minipage}%
    \hspace{5pt}
    \begin{minipage}[t]{.57\linewidth}
        \centering
        \caption{Internal server errors detected over 10 runs.}
        \vspace{-7pt}
        \resizebox{\columnwidth}{!}{%
        \begin{tabular}{lccccc}
        \toprule
        \textbf{Service}          & \textbf{MoRest} & \textbf{RESTler} & \textbf{ARAT-RL} & \textbf{EvoMaster} & \textbf{LlamaRestTest} \\ \midrule
        Features-Service          & 10              & 10              & 10               & 10                & 10                   \\ 
        Genome-Nexus              & 10              & 0               & 10               & 0                 & 10                   \\ 
        Language-Tool             & 0               & 0               & 10               & 10                & 10                   \\ 
        Market                    & 10              & 10              & 10               & 10                & 10                   \\ 
        Person-Controller         & 80              & 80              & 80               & 80                & 80                   \\ 
        Project-Tracking-System                   & 10              & 10              & 10               & 10                & 10                   \\ 
        REST-Countries            & 10              & 10              & 10               & 0                 & 10                   \\ 
        
        User-Management           & 10              & 10              & 10               & 10                & 10                   \\ 
        YouTube                   & 0               & 0               & 0                & 0                 & 0                    \\ 
        FDIC                      & 0               & 0               & 10               & 0                 & 10                   \\ 
        Ohsome                    & 0               & 0               & 0                & 0                 & 36                   \\ 
        Spotify                   & 0               & 0               & 0                & 0                 & 8                    \\ \midrule
        \textbf{Total}              & \textbf{140}    & \textbf{130}    & \textbf{160}     & \textbf{130}      & \textbf{204}         \\ \bottomrule
        \end{tabular}
        }
        \label{tab:internal_errors}
    \end{minipage}
\end{table}

\subsection{Experiment Setup}
\label{sec:experiment_setup}

Our experiments were conducted on an M1 MacBook Pro running Sonoma 14.4.1 with 64GB of memory. We installed all the services from our benchmark suite, along with the REST API testing tools we intended to evaluate. To prevent any dependency conflicts, we restored the database for each service before running each test. Each service and tool was hosted with default settings and tested sequentially to avoid interference. We continuously monitored CPU and memory usage to ensure that resource limitations did not impact the performance of the testing tools. Each tool was run for one hour, a standard time budget used in the area~\cite{kim2022automated,zhang2022open,kim2023reinforcement,liu2022morest,golmohammadi2023survey}, to provide a fair comparison. To minimize the effect of randomness on the results, we repeated each test 10 times and averaged the outcomes.\footnote{For the ablation study (RQ5), we performed only three runs due to its extensive evaluation time.}

For our evaluation, we relied on the same set of REST API testing tools and services used by ARAT-RL~\cite{kim2023reinforcement}. Accordingly, we compared AutoRestTest with ARAT-RL~\cite{kim2023reinforcement}, EvoMaster~\cite{arcuri2019restful}, MoRest~\cite{liu2022morest}, and RESTler~\cite{atlidakis2019restler}. Specifically, we used the latest released version or the latest commit when a release was unavailable: RESTler v9.2.4, EvoMaster v3.0.0, ARAT-RL v0.1, MoRest (obtained directly from the authors).


The ARAT-RL benchmark dataset consists of 10 RESTful services.
In addition to these services, we included services from the RESTGPT study~\cite{kim2023leveraging} due to the availability of a ground truth of extracted rules.
Out of the total 16 services, we excluded SCS and NCS because they were written by EvoMaster's authors, and we aimed to avoid potential bias. We also excluded OCVN due to authentication issues. Lastly, we excluded OMDB, which is a toy online service with only one API operation that all testing tools can process in a second. Ultimately, we used 12 services: Features Service, Language Tool, REST Countries, Genome Nexus, Person Controller, User Management Microservice, Market Service, Project Tracking System, Ohsome,\footnote{Ohsome has an open-source version~\cite{ohsomeopen}, but we used the online version~\cite{ohsomeonline} as the open-source version consistently produced server errors.} YouTube-Mock, and Spotify. Table~\ref{tab:service_lines_of_code} provides details on these services, including the lines of code (for the open-source services) and the number of API operations.


The key metrics for evaluating the effectiveness and fault-detection ability included code coverage, the number of successfully processed operations in the specification, and the number of 500 status codes encountered, which are among the most commonly used metrics~\cite{golmohammadi2023survey}. To measure code coverage, we used JaCoCo~\cite{jacoco}. The number of processed operations and the number of errors detected (500 status codes) were collected using a script from the ARAT-RL repository~\cite{aratartifact}. That script tracks 500 status codes for each operation and removes duplicate 500 errors based on server response messages.

\subsection{RQ1: Comparison with RESTGPT}
\label{rq1}

\begin{table}[t]
\caption{Comparison of semantically valid parameter value generation by LlamaREST-EX, Llama3-8B, and RESTGPT (`FT' indicates fine-tuning and `Q' indicates quantization).}
\vspace{-7pt}
\label{tab:llamarestex}
\centering
\resizebox{.8\columnwidth}{!}{%
\begin{tabular}{lcccccc}
\toprule
\textbf{API} & \multicolumn{4}{c}{\textbf{LlamaREST-EX}} & \textbf{RESTGPT} & \textbf{Llama3-8B} \\
\cmidrule(lr){2-5}
            & \textbf{FT with Q(2-Bit)} & \textbf{FT with Q(4-Bit)} & \textbf{FT with (Q 8-Bit)} & \textbf{FT} & & \\
\midrule
Genome-Nexus    & 10.66\%  & 35.29\% & 35.48\% & 57.02\%  & 36.25\% & 9.00\%\\
Language-Tool   & 22.58\% & 58.82\% & 64.52\% & 69.23\% & 82.98\% & 60.00\% \\
REST-Countries  & 33.33\% & 78.79\% & 75.47\% & 79.22\% & 75.29\% & 46.84\% \\
YouTube         & 14.75\% & 67.27\% & 70.31\% & 73.08\%  & 73.33\% & 21.67\%\\
FDIC            & 21.21\% & 33.33\% & 47.46\% & 54.69\% & 70.42\% & 14.29\% \\
Ohsome          & 27.59\% & 74.14\% & 93.75\% & 92.16\% & 76.83\% & 2.78\%  \\
Spotify         & 73.74\% & 73.81\% & 75.86\% & 81.71\% & 66.67\% & 0.00\%  \\
\midrule
\textbf{Average} & 29.12\% & 60.21\% & 66.12\% & 72.44\% & 68.82\% & 22.94\% \\
\bottomrule
\end{tabular}
}
\end{table}

\begin{table}[t]
\caption{Comparison of inter-parameter dependency rule generation performance among LlamaREST-IPD, RESTGPT, and Llama3-8B (`FT' indicates fine-tuning, `Q' indicates quantization, `O' indicates correct rule generation, and `-' indicates failure to generate the rule).}
\vspace{-7pt}
\label{tab:llamarestipd}
\centering
\resizebox{\columnwidth}{!}{%
\begin{tabular}{llccccccc}
\toprule
\textbf{Service Name} & \textbf{Parameter Name} & \textbf{Rule} & \multicolumn{4}{c}{\textbf{LlamaREST-IPD}} & \textbf{RESTGPT} & \textbf{Llama3-8B} \\
\cmidrule(lr){4-7}
 & & & \textbf{FT with Q(2-Bit)} & \textbf{FT with Q(4-Bit)} & \textbf{FT with Q(8-Bit)} & \textbf{FT} & & \\
\midrule
\multirow{2}{*}{Language-Tool} & data & Or(text,data) & - & - & - & - & O & - \\
 & preferredVariants & IF preferredVariants THEN language==auto & - & - & - & - & - & - \\
\midrule
\multirow{11}{*}{YouTube} & eventType & IF eventType THEN type=video & - & O & O & O & O & O \\
 & forMine & IF forMine==true THEN type=video & - & - & - & O & O & - \\
 & videoLicense & IF videoLicense THEN type=video & - & - & - & O & - & - \\
 & videoType & IF videoType THEN type=video & - & O & O & O & O & O \\
 & videoEmbeddable & IF videoEmbeddable THEN type=video & O & - & O & O & O & - \\
 & videoDimension & IF videoDimension THEN type=video & - & O & O & O & O & - \\
 & location & AllOrNone(this,locationRadius) & - & - & - & - & - & - \\
 & videoDefinition & IF videoDefinition THEN type=video & - & - & - & O & - & - \\
 & relatedToVideoId & IF relatedToVideoId THEN type=video & - & - & - & - & - & - \\
 & videoCaption & IF videoCaption THEN type=video & - & - & O & O & O & - \\
 & forContentOwner & IF forContentOwner THEN onBehalfOfContentOwner & - & - & - & - & - & - \\
 & videoDuration & IF videoDuration THEN type=video & - & O & O & O & O & - \\
 & videoSyndicated & IF videoSyndicated THEN type=video & - & O & O & O & - & - \\
 & videoCategoryId & IF videoCategoryId THEN type=video & - & O & - & O & - & - \\
\midrule
Spotify & collaborative & IF collaborative==true THEN public==false & - & - & - & O & O & - \\
\midrule

\textbf{Total} & & 18 & 1 & 6 & 7 & 12 & 9 & 2 \\
\bottomrule
\end{tabular}
}
\end{table}
To assess the effectiveness of LlamaREST-IPD in identifying IPD rules and LlamaREST-EX in generating semantically valid testing inputs, we compared both models with RESTGPT, a state-of-the-art tool for enhancing specifications.  We also evaluated Llama3-8B, the base model for both LlamaREST-IPD and LlamaREST-EX, along with various quantization variants (2-bit, 4-bit, and 8-bit). To measure effectiveness, we used as ground truth, the RESTGPT dataset of extracted rules, including IPDs and example values~\cite{kim2023leveraging}. The RESTGPT dataset contains rules manually extracted from the OpenAPI specifications of nine services, but IPD and value constraint rules occur for only three and seven of the services, respectively; for the other services, these rules were not described in their OpenAPI specifications. Thus, our evaluation of value generation is based on seven of the services from the RESTGPT benchmark, whereas the evaluation of IPD rule extraction is based on three of the services.


Table~\ref{tab:llamarestex} presents the results on the precision of semantically valid value generation for LlamaREST-EX, RESTGPT, and Llama3-8B for seven services. Table~\ref{tab:llamarestipd} further narrows the focus to three of these services, presenting the available IPD rules identified through parameter descriptions. 

We evaluated value generation based on semantic validity, as proposed by a recent study and adopted by RESTGPT: a semantically valid input must be coherent with the API domain~\cite{alonso2022arte}. For instance, ``Berlin'' is a valid input for the parameter \texttt{\small addressLocality} in DHL's API, whereas ``dog'' is not. For RESTGPT, we used the generated values from its artifact; for Llama3-8B, we applied the same prompts used by RESTGPT. Precision was determined by two external developers independently assessing whether the generated values were semantically valid, with any discrepancies resolved through discussion to reach a consensus.

Llama3-8B, although competitive on some APIs, generally performs poorly, with an average accuracy of just 24.94\%. It struggles significantly on APIs such as Spotify (0\%) and Ohsome (2.78\%), underscoring its limitations in generating semantically valid values. However, fine-tuning the model, as demonstrated with LlamaREST-EX, leads to significant performance improvement.

LlamaREST-EX, especially its FT version, shows substantial gains over Llama3-8B, achieving an average accuracy of 72.44\%. For example, on the Ohsome API, the FT version of LlamaREST-EX reaches 92.16\%, outperforming RESTGPT’s 76.83\%; similarly, on Genome-Nexus, it scores 57.02\%, achieving considerably higher precision than RESTGPT (36.25\%). Overall, LlamaREST-EX with fine-tuning outperforms RESTGPT, which achieved 68.82\% accuracy, compared to LlamaREST-EX’s 72.44\%. Even the quantized versions of LlamaREST-EX (4-bit and 8-bit) show reasonable performance, with 60.21\% and 66.12\% precision, respectively.

LlamaREST-EX generally performs well, but there are two cases where it is less effective than RESTGPT: Language-Tool and FDIC. These results highlight some limitations of our approach and provide insights for future improvements. For Language-Tool, the low performance is primarily due to LlamaREST-EX's difficulty in handling complex parameter constraints, particularly for region-specific formats. For example, with the \texttt{\small altLanguage} parameter, LlamaREST-EX output ``en'' instead of the correct ``en-GB''. RESTGPT, leveraging its larger model size, was able to correctly interpret the format specification from the parameter description. In the case of FDIC, the lower performance came from the specialized nature of the API, which requires more domain-specific financial knowledge that our fine-tuning process did not fully capture.

These cases highlight the trade-off between model size and performance. Although our approach with LlamaRestTest demonstrates that fine-tuning a smaller, open-source model can significantly enhance task-specific effectiveness and make it competitive with much larger models in many cases, there are still scenarios where larger models can have an advantage in handling very specific or complex constraints.




Table~\ref{tab:llamarestipd} presents the IPD rules identified through parameter descriptions for three services among the seven services in Table~\ref{tab:llamarestex}. Llama3-8B demonstrates limited capability in detecting IPD rules, identifying only two rules across all services. LlamaREST-IPD, particularly in its fine-tuned version, significantly outperforms both Llama3-8B and RESTGPT. The FT variant of LlamaREST-IPD identifies 12 correct IPD rules, compared to nine by RESTGPT, highlighting its ability to capture more complex relationships. Even the quantized versions of LlamaREST-IPD (4-bit and 8-bit) demonstrate consistent improvements in performance, but the 2-bit version has considerable performance degradation, identifying only one IPD rule correctly.

\begin{rqanswer}
Fine-tuning significantly improves the accuracy of semantically valid input generation and IPD rule detection, outperforming both the base Llama3-8b model and the state-of-the-art RESTGPT tool. Moreover, with 8-bit and 4-bit quantization, the fine-tuned model remains competitive with RESTGPT, while still performing considerably better than the base model.
\end{rqanswer}

\begin{figure}[t]
\centering
\includegraphics[width=\columnwidth]{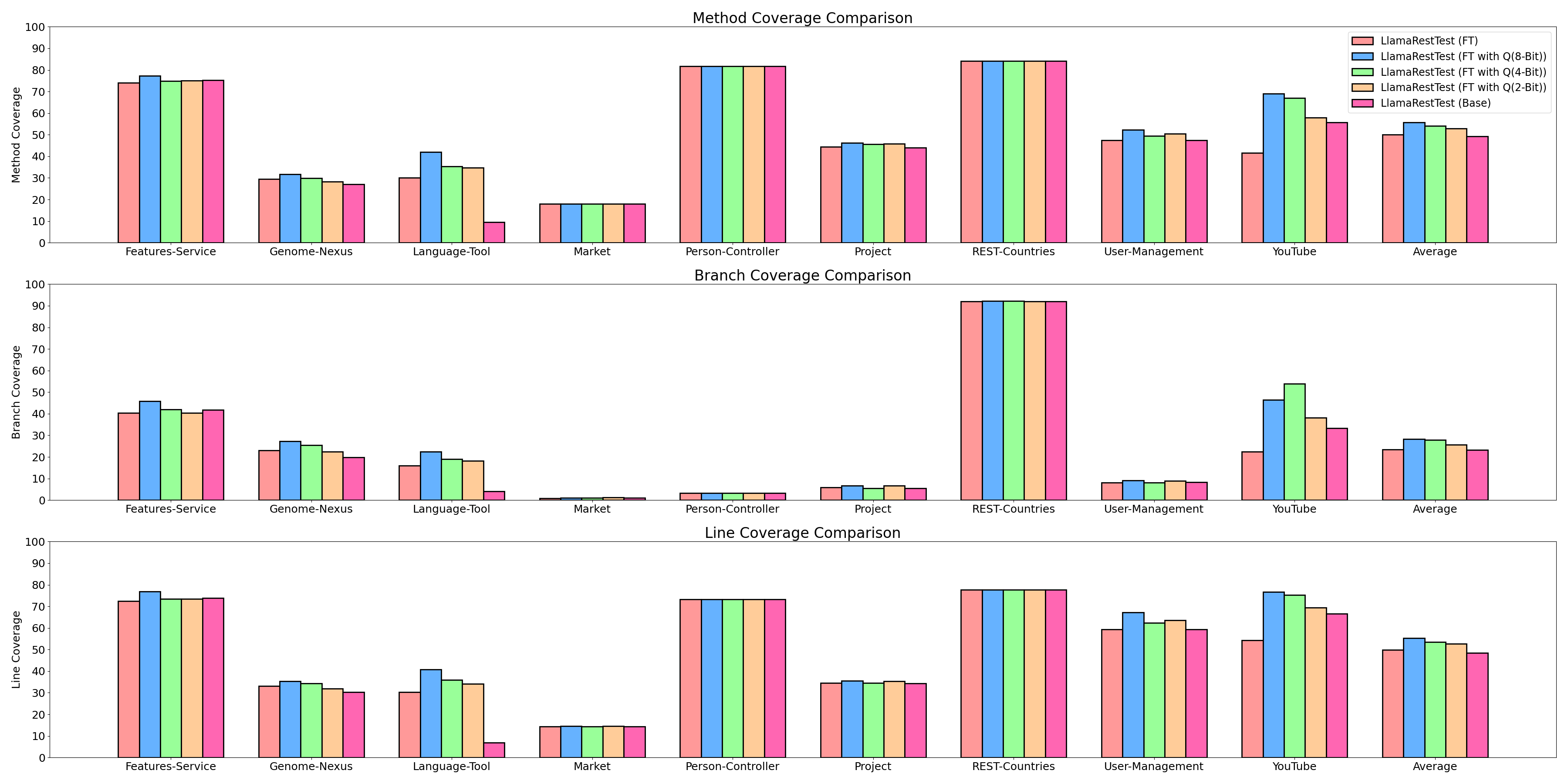}
\caption{Code coverage achieved on 10 runs for open-source services using LlamaRestTest configured with the base Llama3-8B model, the fine-tuned model (FT), and the quantized models (Q).}
\label{fig:coverage}
\end{figure}


\subsection{RQ2: Impact of Fine-Tuning and Quantization on REST API Testing}
\label{rq2}

\begin{table}[t]
\centering

\begin{minipage}{0.48\textwidth}
\caption{Average number of operations processed in 10 runs for closed-source services using LlamaRestTest configured with Llama3-8B (Base), fine-tuning (FT), and quantization (Q).}
\vspace{-7pt}
\label{tab:processed_op}
\centering
\resizebox{\textwidth}{!}{
\begin{tabular}{lccccc}
\toprule
\textbf{Service} & \textbf{FT with 8-Bit Q} & \textbf{FT with 4-Bit Q} & \textbf{FT with 2-Bit Q} & \textbf{FT} & \textbf{Base} \\ \midrule
FDIC     & 6.0  & 6.0  & 6.0  & 6.0  & 6.0  \\ 
Ohsome   & 24.5 & 21.4 & 13.2 & 12.1 & 11.2 \\
Spotify  & 7.0  & 7.0  & 6.2  & 6.4  & 5.8  \\ 
 \midrule
\textbf{Total}     & \textbf{37.5} & \textbf{34.4} & \textbf{25.4} & \textbf{24.5} & \textbf{23.0} \\ \bottomrule
\end{tabular}
}
\end{minipage}%
\hfill
\begin{minipage}{0.48\textwidth}
\caption{Average number of operations processed for closed-source services by different testing tools (LlamaRestTest configured with 8-bit quantization).}
\label{tab:op_comparison}
\centering
\resizebox{\textwidth}{!}{
\begin{tabular}{lccccc}
\toprule
\textbf{Service} & \textbf{MoRest} & \textbf{RESTler} & \textbf{ARAT-RL} & \textbf{EvoMaster} & \textbf{LlamaRestTest} \\ \midrule
FDIC     & 6.0  & 6.0  & 6.0  & 6.0  & 6.0  \\
Ohsome   & 0.0  & 0.0  & 0.0  & 0.0  & 24.5 \\
Spotify  & 3.9  & 4.6  & 5.6  & 5.2  & 7.0  \\ 
 \midrule
\textbf{Total}     & \textbf{9.9}  & \textbf{10.6} & \textbf{11.6} & \textbf{11.2} & \textbf{37.5} \\ \bottomrule
\end{tabular}
}
\end{minipage}

\end{table}

To evaluate the impact of fine-tuning and various quantization levels on REST API testing, we assessed LlamaRestTest under five configurations: the base Llama3-8B model, the fine-tuned version (at full precision), and the quantized versions at 8-bit, 4-bit, and 2-bit quantization. We evaluated the effectiveness of these configurations in terms of code coverage achieved for the open-source services and operations processed for the closed-source services in our benchmark (Table~\ref{tab:service_lines_of_code}).

Figure~\ref{fig:coverage} presents a comparison of method, branch, and line coverage across multiple services for the five LlamaRestTest configurations: the fine-tuned version, the fine-tuned versions with 8-bit, 4-bit, and 2-bit quantization, and the base model. Each panel represents one of the coverage metrics---method, branch, or line---for the nine open-source services in our benchmark.  The figure also shows the average coverage across services for each configuration. The fine-tuned and quantized versions generally exhibited higher coverage compared to the base model, with significant gains observed for the Language-Tool and YouTube services in particular.

The results show that fine-tuning alone produced modest improvements over the base model. The fine-tuned model achieved average coverage rates of 50.11\% for method coverage (compared to 49.21\% for base), 23.55\% for branch coverage (compared to 23.24\% for base), and 49.91\% for line coverage (compared to 48.51\% for base). This represents improvements of approximately 0.9, 0.3, and 1.4 percentage points respectively, demonstrating that fine-tuning provides an observable but limited enhancement to testing capability.

On combining fine-tuning with quantization, we observed substantially greater improvements. LlamaRestTest with 8-bit quantization exhibited 6.6 percentage points improvement in method coverage compared to the base model, reaching 55.8\% coverage on average. The 4-bit quantized version resulted in a coverage gain of 4.8 percentage points (54.0\% method coverage), while the 2-bit quantized version achieved 3.7 more percentage points (52.9\% method coverage). The trends are similar for branch and line coverage, with the quantized versions performing better than the fine-tuned model in gains over the base model. In terms of branch coverage, the 8-bit, 4-bit, and 2-bit quantized versions showed increases of 5.1, 4.7, and 2.5 percentage points, respectively, over the base model; likewise, for line coverage, the gains of these quantized versions were 6.8, 5.0, and 4.1 percentage points, respectively.

Table~\ref{tab:processed_op} shows the number of successfully processed operations for the closed-source services in our benchmark under the five LlamaRestTest configurations. LlamaRestTest with fine-tuning alone processed 24.5 operations, which is 1.5 operations more than the base model's 23 operations. More significantly, LlamaRestTest with 8-bit quantization processed 37.5 operations (14.5 operations more than the base model's 23), while the 4-bit quantized version processed 34.4 operations (11.4 operations more than the base model). Finally, the 2-bit version was also more effective than the base model, processing 25.4 operations (2.4 operations more than the base model).

These results demonstrate that fine-tuning alone provides modest benefits, but the combination of fine-tuning with appropriate quantization delivers significantly better performance. There are two main contributing factors for this phenomenon. First, the faster inference times of quantized models enable more test generation within the same testing budget, processing for instance 22--81\% more inference tasks in the same time budget~\cite{10.5555/3600270.3602468}. For example, for IPD rule extraction (RQ1), the fine-tuned model required 48.9 seconds per IPD on average, whereas the 8-bit quantized version reduced the inference time to 36.9 seconds. The 4-bit and 2-bit quantized models further reduced the time to 26.2 and 26.1 seconds, respectively. The smaller models benefit from faster inference, which enhances the effectiveness of REST API testing via more efficient processing of server response messages.
Second, quantization introduces a regularization effect that may help the model generalize better with various answers rather than memorizing specific examples~\cite{gholami2022survey}.

Although quantized models generally performed well, we observed scenarios where the 2-bit quantized model underperformed compared to the non-quantized model. Specifically, for services, such as Genome-Nexus and Spotify, that rely heavily on domain-specific knowledge, the 2-bit quantized model often generated incorrect outputs, leading to lower code coverage. This suggests that the extreme compression in the 2-bit model may result in a loss of precision that affects performance in complex, domain-specific scenarios. However, the 4-bit and 8-bit quantized models consistently outperformed the non-quantized models, even in these challenging cases. This indicates that these quantization levels strike a good balance between model compression and maintaining the necessary precision for effective API testing across various domains.

\begin{rqanswer}
LlamaRestTest with fine-tuning and 8-bit quantization achieved the highest code and operation coverage, processing 37.5 operations while achieving improvements of 5.5, 5.1, and 6.8 percentage points in method, branch, and line coverage, respectively, over the base model, thus showing a good balance between efficiency and effectiveness. Fine-tuning alone provided modest gains over the base model. While 4-bit and 8-bit quantized models consistently outperformed non-quantized versions, the 2-bit model showed limitations in complex, domain-specific scenarios, highlighting the trade-off between model compression and precision for REST API testing.
\end{rqanswer}

\subsection{RQ3: Comparison with State-of-the-art REST API Testing Tools}

\begin{figure}[t]
\centering
\includegraphics[width=\columnwidth]{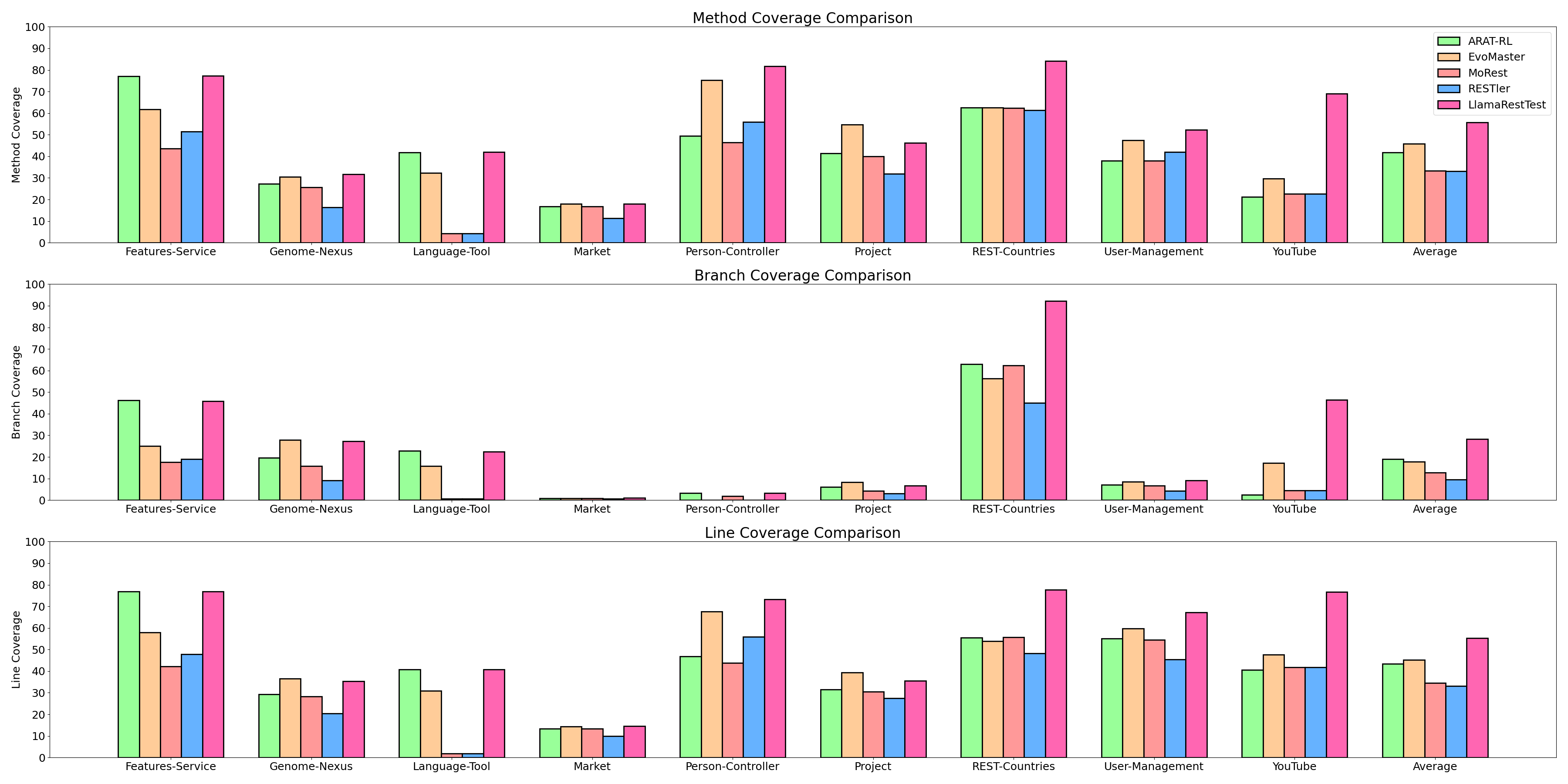}
\caption{Code coverage achieved on 10 runs for open-source services with MoRest, RESTler, ARAT-RL, EvoMaster, and LlamaRestTest (fine-tuned with 8-bit quantization).}
\Description{Code Coverage Achieved Across 10 Runs for Open Source Services with MoRest, RESTler, ARAT-RL, EvoMaster, and LlamaRestTest (Fine-Tuning with 8-Bit Quantized)}
\label{fig:comparison}
\end{figure}

Next, we compared LlamaRestTest against four state-of-the-art REST API testing tools in terms of code coverage (for the open-source services) and operation coverage (for the closed-source services). For this study, the testing tools used enhanced versions of OpenAPI specifications of the services generated by RESTGPT~\cite{kim2023leveraging}.

Figure~\ref{fig:comparison} presents the code coverage achieved by MoRest, RESTler, ARAT-RL, EvoMaster, and LlamaRestTest on the open-source services in our benchmark. The results show that LlamaRestTest consistently outperforms all other tools in method, branch, and line coverage. For method coverage, LlamaRestTest achieved 55.8\%, exceeding MoRest (33.3\%) by 22.5 percentage points, RESTler (33.1\%) by 22.7 percentage points, ARAT-RL (41.7\%) by 14.1 percentage points, and EvoMaster (45.8\%) by 10 percentage points.
In terms of branches, LlamaRestTest reached 28.3\% coverage, achieving 15.6 increase over MoRest, 18.7 increase over RESTler, 9.2 increase over ARAT-RL, and 10.5 increase over EvoMaster.
Finally, for line coverage, LlamaRestTest achieved 55.3\% coverage, again outperforming MoRest (+20.7), RESTler (+22.1), and ARAT-RL (+11.9), and EvoMaster (+10) considerably.

Table~\ref{tab:op_comparison} presents results on operation coverage, showing the average number of operations processed by each tool for the closed-source services. Like the results on code coverage, LlamaRestTest outperformed the other tools by a wide margin, successfully processing 37.5 operations overall. This represents improvements of 27.6 operations over MoRest (9.9), 26.9 operations over RESTler (10.6), 25.9 operations over ARAT-RL (11.6), and 26.3 operations over EvoMaster (11.2).

LlamaRestTest's effective performance comes from its ability to analyze server responses, which allows it to detect errors that other tools can miss. For example, in testing Ohsome and Spotify APIs, LlamaRestTest was able to process more operations by utilizing information from server messages. In the case of Ohsome, where the server returned a message indicating that users must specify only one of three parameters, LlamaRestTest used this information to test the operations that the other tools missed. 
Conversely, LlamaRestTest may be less effective on services that provide minimal or poor server feedback. In such cases, LlamaRestTest could spend time analyzing response messages that do not contain useful information for refining requests, potentially reducing its efficiency.

These results demonstrate that leveraging server response feedback with LlamaREST-IPD and LlamaREST-EX significantly improves testing effectiveness compared to current state-of-the-art tools, even when using enhanced specifications generated by RESTGPT. The ability to interpret and utilize server responses gives LlamaRestTest an edge in most scenarios, particularly those with informative server feedback.

\begin{rqanswer}
LlamaRestTest significantly outperforms other API testing tools, achieving 10.0--22.7 percentage points higher method coverage, 9.2--18.7 percentage points higher branch coverage, 10.0--22.1 percentage points higher line coverage, and processing 2.5x--3x more operations than competing tools. It performs particularly well in scenarios with informative server feedback, but may be less effective---performing similarly to other tools---on APIs with response messages that lack useful information.
\end{rqanswer}

\subsection{RQ4: Fault-Detection Capability}

Table~\ref{tab:internal_errors} presents our results on fault-detection capability of five REST API testing tools—MoRest, RESTler, ARAT-RL, EvoMaster, and LlamaRestTest—measured as the number of 500 internal server errors detected across all services in our benchmark. To measure unique server errors, we used the error-counting script provided in the ARAT-RL artifact~\cite{aratartifact}, which automatically counts the 500 errors in the server log and discards duplicate entries. 

LlamaRestTest demonstrated the highest fault-detection ability, discovering 204 faults in 10 runs on the services in our benchmark. This represents 64 more faults detected than MoRest (140), 74 more than RESTler (130), 44 more than ARAT-RL (160), and 74 more than EvoMaster (130). The results clearly show that LlamaRestTest outperforms all other tools, detecting more faults across diverse services.

The most notable detected errors were in the Ohsome and Spotify services. For Ohsome, LlamaRestTest was the only tool to detect faults, identifying 36 errors where all other tools failed. Similarly, LlamaRestTest was the only tool to find faults in Spotify, detecting 8 errors where all others found none. Both if these instances involve processing of complex scenarios. For example, the \textsc{get} \texttt{\small /playlists/{playlist\_id}/tracks} operation in Spotify's API requires specific knowledge of how Spotify generates \texttt{\small playlist\_id}. Spotify generates IDs for playlists that are typically 22 characters long, with constraints on the permitted characters and patterns. While many tools fail to generate valid IDs, LlamaREST-EX, due to its fine-tuning, accurately retrieved and generated valid Spotify playlist IDs. As a result, LlamaRestTest was able to successfully execute the \textsc{get} \texttt{\small /playlists/{playlist\_id}/tracks} operation, triggering a chain of interactions across the service. For instance, after retrieving the International Standard Recording Code (ISRC) from the playlist's tracks, the mutation process randomly selects an ISRC to use as \texttt{\small user\_id} in the subsequent \textsc{get} \texttt{\small /users/{user\_id}/playlists} operation. This sequence exposes a hidden dependency conflict, which causes an Internal Server Error.

\begin{rqanswer}
LlamaRestTest outperforms state-of-the-art REST API testing tools in fault detection by triggering 204 Internal Server Error (500) responses over 10 runs, thus identifying 44 to 74 more errors than the other tools, which illustrates its superior ability to detect internal server errors.
\end{rqanswer}

\subsection{RQ5: Ablation Study}

\begin{table}[t]
\caption{Ablation study on the impact of key components of LlamaRestTest on coverage achieved.}
\vspace{-7pt}
\centering
\label{tab:ablation}
\resizebox{0.5\columnwidth}{!}{%
\begin{tabular}{lccc}
\toprule
 & Method & Branch & Line \\
\midrule
LlamaRestTest & 55.8\% & 28.3\% & 55.3\% \\
1. Without Parameter Description & 49.9\% (-5.9\%) & 26.9\% (-1.4\%) & 50.8\% (-4.5\%) \\
2. Without Server Response & 49.7\% (-6.1\%) & 23.9\% (-4.4\%) & 49.0\% (-6.3\%) \\
3. Without LlamaREST-IPD & 46.6\% (-9.2\%) & 22.1\% (-6.2\%) & 46.9\% (-8.4\%) \\
4. Without LlamaREST-EX & 52.1\% (-3.7\%) & 23.8\% (-4.5\%) & 52.3\% (-3.0\%) \\
\bottomrule
\end{tabular}
}
\end{table}

To assess the contribution of the key components of LlamaRestTest, we conducted an ablation study by systematically removing specific features and measuring the resulting effect on method, branch, and line coverage across three runs. In particular, we disabled the following features individually to study their effects on coverage: analysis of parameter descriptions, analysis of server response messages, LlamaREST-IPD, and LlamaREST-EX. Table~\ref{tab:ablation} presents the code coverage achieved by LlamaRestTest with and without these components, thus showing how much each feature contributes to its overall performance.

The full LlamaRestTest system achieved method coverage of 55.8\%, branch coverage of 28.3\%, and line coverage of 55.3\%. Removing the parameter description feature resulted in a noticeable drop, with method coverage decreasing by 5.9 percentage points, branch coverage by 1.4 percentage points, and line coverage by 4.5 percentage points. This suggests that including parameter descriptions improves test case generation by providing relevant information for generation of valid inputs. Interestingly, excluding the server response feature caused an even more significant drop, particularly in method coverage, which fell by 6.1 percentage points, followed by branch and line coverage, which decreased by 4.4 and 6.3 percentage points, respectively. While previous studies in this area have focused primarily on enhancing specifications~\cite{kim2023enhancing,kim2023leveraging}, these results highlight the critical role of server feedback in guiding test case generation and increasing coverage via analysis of dynamic behaviors at runtime.

The removal of LlamaREST-IPD had the most substantial impact on performance, reducing method coverage by 9.2 percentage points, branch coverage by 6.2 percentage points, and line coverage by 8.4 percentage points. This suggests that modeling inter-parameter dependencies is crucial for generating more effective test cases that cover complex interactions between parameters. Finally, excluding LlamaREST-EX resulted in moderate reductions in coverage, with method, branch, and line coverage decreasing by 3.7, 4.5, and 3.0 percentage points, respectively. But LlamaREST-EX still contributes meaningfully to overall test effectiveness, particularly in terms of branch coverage.

\begin{rqanswer}
The ablation on key components of LlamaRestTest shows that, while each component contributes to the overall performance of LlamaRestTest, server response analysis and inter-parameter dependency modeling are the more critical components, with their removal causing larger drops in code coverage than the other components. 
\end{rqanswer}

\subsection{Threats to Validity}
\label{subsec:threats_to_validity}

Like any empirical study, there are potential threats to the validity of our results.  Our selection of REST APIs may not be representative and affect external validity. Although we used a diverse set of 12 real-world services, including services that have been used in prior evaluations of REST API testing tools, LlamaRestTest may perform differently on other benchmarks containing services with different characteristics. For instance, our reliance on specific fine-tuning datasets, such as Martin-Lopez’s IPD database and APIs-guru, may affect the performance of our fine-tuned models on other services.

There are also possible threats related to the experiment setup that could affect the accuracy of our results. First, we only compared with a limited number of tools. To address this threat, we considered the best-performing state-of-the-art REST API testing tools available at the time of this study. Second, the settings used for the testing tools and services might not be optimal. To mitigate this threat, we used the same settings used in a previous study~\cite{kim2023reinforcement} and hosted each testing tool and service on separate machines to prevent interference. Furthermore, to account for randomness in tool results, we ran each tool 10 times on each service (this was done for all studies except the ablation study). Third, our use of automated scripts to measure fault detection and code coverage could introduce inaccuracies due to bugs in the scripts. However, we used scripts that have been used in prior studies~\cite{aratartifact} and double-checked them, in addition to manually spot-checking some of the results. 
Finally, LlamaRestTest's integration with ARAT-RL creates tool dependency, in that potential issues in ARAT-RL could affect LlamaRestTest's performance. Also in this case, we thoroughly tested and spot-checked implementation and results. We also made LlamaRestTest and the experiment dataset available in our artifact~\cite{artifact}, which allows other researchers to further validate and extend our evaluation.

\section{Related Work}
\label{sec:related}

\textbf{Automated REST API Testing.}
Automated testing for REST APIs has seen various strategies. White-box and black-box approaches, such as those in EvoMaster, leverage evolutionary algorithms to refine test cases dynamically, with a focus on internal server errors (500 status code) detection~\cite{arcuri2019restful}. Black-box methodologies, exemplified by RESTler~\cite{atlidakis2019restler}, generate stateful tests by deducing producer-consumer dependencies, aiming at detecting server failures. RestTestGen~\cite{Corradini2022} and similar tools exploit data dependencies and employ oracles for validating status codes and response schemas. MoRest~\cite{liu2022morest} focuses on model-based testing to simulate user interactions and test case generation, while RestCT~\cite{wu2022combinatorial} employs combinatorial testing techniques to systematically explore parameter value combinations and their effects on API behavior. ARAT-RL~\cite{kim2023reinforcement} introduces a reinforcement learning approach, dynamically refining API testing strategies by adapting to real-time feedback. Techniques like QuickREST~\cite{karlsson2020quickrest}, Schemathesis~\cite{zac2022schemathesis}, and RESTest~\cite{martin2021restest} use property-based testing and various oracles to ensure response compliance with OpenAPI or GraphQL specifications. Tools such as Dredd~\cite{dredd}, fuzz-lightyear~\cite{fuzz-lightyear}, and Tcases~\cite{tcases} offer diverse testing capabilities, from comparing expected responses to detecting vulnerabilities and validating Swagger-based specifications. 
More recently, two additional REST API testing techniques were proposed: DeepREST~\cite{corradini2024deeprest} and AutoRestTest~\cite{kim2024multi}. DeepREST uses deep reinforcement learning to discover implicit API constraints, employing a single agent that learns operation orderings through a reward mechanism. AutoRestTest uses LLMs to generate parameter values while relying on the semantic property dependency graph for efficient dependency identification and multi-agent reinforcement learning for dynamic optimization across all testing steps. In contrast, LlamaRestTest employs fine-tuned LLMs for detecting IPDs and generating semantically valid input values, as well as leveraging reinforcement learning via its integration with ARAT-RL. It is worth noting that because DeepREST and AutoRestTest were developed concurrently with this work, we were unable to include them in our empirical comparisons.

\textbf{LLM-based Test Generation.} There is a growing body of work on leveraging LLMs for test generation~\cite{wang2024survey}, much of it focusing on unit testing. For example, AthenaTest~\cite{tufano2020unit} uses a fine-tuned model for unit test generation, whereas TestPilot~\cite{schafer2023empirical} synthesizes JavaScript and TypeScript tests. ASTER~\cite{pan2025asternaturalmultilanguageunit} implements an LLM-based pipeline for generating unit test for Java and Python. Unlike these techniques, LlamaRestTest focuses on REST API testing, employing two fined-tuned models for specific tasks together with analysis of dynamic server feedback.

\textbf{LLM-based REST API Analysis.} Recent work has also applied LLMs for REST API specification inference and API invocation. For instance, RESTSpecIT~\cite{decrop2024you} demonstrated capabilities for automatically inferring API specifications and conducting black-box testing with minimal user intervention.
NESTFUL~\cite{basu2024nestful} created a benchmark for evaluating how LLMs handle complex nested API calls, highlighting current limitations in processing interconnected API operations.  RestGPT~\cite{song2023restgpt} introduced a coarse-to-fine online planning approach that enhances task breakdown and API selection processes. 
In contrast, LlamaRestTest leverages fine-tuned LLMs, server response messages, and reinforcement learning for improving REST API testing.



\section{Conclusion and Future Work}
\label{sec:conclusion}

We introduced LlamaRestTest, a new REST API testing approach that leverages the capabilities of large language models through fine-tuning and quantization to improve testing effectiveness and efficiency. LlamaRestTest incorporates two key components, LlamaREST-IPD and LlamaREST-EX, which focus on inter-parameter dependency detection and realistic parameter value generation. Our evaluation, performed on 12 real-world REST services, shows that LlamaRestTest can outperform state-of-the-art tools such as RESTler, EvoMaster, MoRest, and ARAT-RL, achieving significantly higher code coverage and processing more service operations than those tools. Additionally, our ablation study highlights the importance of each component, with LlamaREST-IPD and server response messages being critical to achieving the best results. Our results also show that fine-tuning significantly enhances LlamaRestTest’s performance, even outperforming larger models such as GPT. Quantization further improves the model's efficiency without substantial loss in accuracy, making LlamaRestTest both powerful and practical for real-world use. In future work, we will explore ways to expand our approach's capabilities to handle a wide variety of API types and refine its dynamic interaction with server responses for even more effective test input generation.



\section*{Acknowledgments}
\begin{small}
  This work was partially supported by 
  NSF, under grant CCF-0725202 and
  DOE, under contract DE-FOA-0002460,
  and gifts from Facebook, Google, IBM Research, and Microsoft Research.
\end{small}
\balance
\bibliographystyle{ACM-Reference-Format}
\bibliography{bib}

\end{document}